\documentclass{article}%
\usepackage{amssymb}
\usepackage{graphicx}
\usepackage{amsmath}%
\usepackage{amsfonts}
\begin{document}

\title{W-T identities and a candidate ``droplet'' Lagrangean for the Ising Spin Glass}
\author{Cirano De Dominicis\\Service de Physique Th\'{e}orique\\CEA/Saclay - DSM/SPhT\\F-91191 Gif sur Yvette Cedex, France}
\maketitle
\date{}

\begin{abstract}
In search for a microscopic description of \textquotedblleft
droplet-like\textquotedblright\ properties for the Ising spin glass (single
component order parameter, zero modes i.e. correlation functions vanishing at
infinity) we reconsider the two-packet model of Bray and Moore, which is
effectively Replica-Symmetric and enjoys zero modes but only up to one-loop.
We show how an appropriate change in the limits of the basic parameters of the
model (packet size and replica number) allows for a derivation of
Ward--Takahashi (WT) identities, thus ensuring the existence of zero modes to
all orders and opening the way for a Lagrangean formulation of a
\textquotedblleft droplet-like\textquotedblright\ field theory for the Ising
spin glass.

\end{abstract}

Spin glass theory has presented so far the schizophrenic aspect of two
conflicting approaches difficult to reconcile. The so-called mean-field like
approach developed around Parisi \cite{Parisi1}-\cite{Parisi3} solution of the
Sherrington-Kirkpatrick \cite{SK} mean-field model uses standard field theory
(mean-field, loops, renormalisation group, WT identities, ...) on a Lagrangean
built with replicated fields. It is a \textquotedblleft
microscopic\textquotedblright\ theory for the spin glass, in the same sense as
the $\phi^{4}$ theory for the ferromagnet. The alternative droplet-like theory
of Fisher and Huse \cite{FisherHuse} and Bray and Moore \cite{BMdroplets},
despite abundant results, does not appeal to such a Lagrangean field theory
starting point and is far removed from \textquotedblleft
microscopic\textquotedblright\ description. So there is a strong motivation to
search better into the replicated $\phi^{3}$ Lagrangean and examine whether
the characteristic features for a \textquotedblleft droplet\textquotedblright%
\ like theory could fit in.

If one were to put up one, it would have the features of a theory with an
effective Replica-Symmetric order parameter (a ferromagnetic in disguise). It
would also possess zero modes to allow for an algebraic decrease at large
distances of its correlations. But such features were indeed present, years
before the birth of droplet theory, in an ansatz proposed by Bray and Moore
\cite{BM79} (BM), were replica symmetry was broken into \textquotedblleft
two-packets\textquotedblright, and restored at the very end. From their
results, calculated at one-loop, it could be checked that both features
mentioned above (no RSB, zero modes) were present. However there was no
guarantee that the zero modes would survive beyond one loop (indeed they would
not and it was suggested that a generalization to multi-packet could help zero
modes to remain massless).

In this note we wish to take a new look at the \textquotedblleft
two-packets\textquotedblright\ RSB, as BM applied it in \cite{BM79} to a spin
glass Lagrangean with cubic coupling. We wish first to understand why this RSB
did not give rise to WT identities that would have protected the zero modes to
all orders. Indeed such identities have been derived \cite{WTidentity} in the
framework of a Lagrangean field theory for systems with R steps of RSB (with
in the end the Parisi limit $R\rightarrow\infty$). One essential ingredient in
the proof is the selection of \textquotedblleft infinitesimal
permutations\textquotedblright, in fact infinitesimal like $1/R$. In the
two-packet theory, one packet has $m$ replicas ($a$ $b$ $c$ ...) and the other
$n-m$ ($\alpha$ $\beta$ $\gamma$ $\ldots$). In the end $n$ is set to zero (as
it should for replicas) and $m$ sent to infinity. Infinitesimal permutations
are then easily selected (they are associated with transverse generators) and
they go to zero with $1/m$. So following the same steps as in
\cite{WTidentity}, we identify below where the model fails to yield WT
identities. Bringing the appropriate change allows then WT identities to be
established, thereby giving life to a \textquotedblleft
droplet\textquotedblright\ Lagrangean field theory for the spin glass.

\section{Framework for WT identities:}

We start with a permutation invariant free-energy functional%
\begin{equation}
F\left\{  Q_{A,B}\right\}  =F\left\{  Q_{PA,PB}\right\} \tag{1}\label{001}%
\end{equation}
$A$ is a replica index that could have been denoted $\left(  i,a\right)  $
$i=1 $ or $2$, but is more economically replaced by $a$ or $\alpha$, the
roman-greek notation of BM. The order parameter $Q_{AB}$ is given by the
stationarity\ condition on (\ref{001}) and as in \cite{BM79}, at mean-field
level one has%
\begin{align}
Q_{ab}  & =Q_{1}=Q\frac{m-n}{m-n/2}\tag{2a}\label{002a}\\
Q_{\alpha\beta}  & =Q_{2}=Q\frac{m}{m-n/2}\tag{2b}\label{002b}\\
Q_{a\alpha}  & =Q_{0}=Q\frac{m}{m-n/2}\left[  1-\frac{n}{m}+\frac{n}{m^{2}%
}\right]  ^{1/2}\tag{2c}\label{002c}%
\end{align}
On the other hand, from the definition of Legendre transform one
has\footnote{For reasons of convenience we have changed the notation of BM
with their $Q_{3}\rightarrow Q_{1}$, $Q_{2}\rightarrow Q_{0}$, $Q_{1}%
\rightarrow Q_{2}$}%
\begin{equation}
W\left\{  H_{AB}\right\}  +F\left\{  Q_{AB}\right\}  =\sum_{\left(  AB\right)
}H_{AB}Q_{AB}\tag{3}\label{003}%
\end{equation}
where $H_{AB}$ is an external (unphysical) conjugate source, and hence%
\begin{equation}
\frac{\partial F\left\{  Q_{CD}\right\}  }{\partial Q_{AB}}=H_{AB}\text{
,}\tag{4}\label{004}%
\end{equation}
yielding stationarity when the source is set to zero.

The invariance under permutation then writes%
\begin{equation}
\frac{\partial F}{\partial Q_{AB}}\left\{  Q_{CD}+\delta Q_{CD}\right\}
=H_{AB}+\delta H_{AB}\tag{5}\label{005}%
\end{equation}
where%
\begin{equation}
Q_{PA;PB}=Q_{AB}+\delta Q_{AB}\tag{6}\label{006}%
\end{equation}
Now we have to make a choice for $P$ (a choice amounting to use transverse
generators). Just like in \cite{WTidentity} we can divide packet one into
$\frac{m}{p}$ equal bunches of $p$ roman replicas and packet two into
$\frac{n-m}{p}$ equal bunches of greek replicas. The permutation chosen is for
example, exchanging the first bunch of $p$ roman replicas ($a_{1}$ $b_{1}$
$c_{1}$ ...) with the first bunch of $p$ greek replicas ($\alpha_{1}$
$\beta_{1}$ $\gamma_{1}$ ...). In the following we keep $a$ $b$ $c$ ... or
$\alpha$ $\beta$ $\gamma$ ... for replicas that do no belong to the exchanged
first bunches.

Let us now look at the effect of such a permutation $P$ by computing $\delta
Q_{CD}$ (or $\delta H_{AB}$). Clearly one has%
\begin{align}
\delta Q_{ab}  & =\delta Q_{\alpha\beta}=\delta Q_{a\alpha}=\delta
Q_{a_{1}\alpha_{1}}=0\tag{7a}\label{007a}\\
\delta Q_{a_{1}b_{1}}  & =-\delta Q_{\alpha_{1}\beta_{1}}=Q_{2}-Q_{1}%
\equiv\delta Q_{0}\tag{7b}\label{007b}\\
\delta Q_{a_{1}\alpha}  & =-\delta Q_{\alpha_{1}\alpha}=Q_{2}-Q_{0}%
\equiv\delta Q_{2}\tag{7c}\label{007c}\\
\delta Q_{a\alpha_{1}}  & =-\delta Q_{a_{1}a}=Q_{1}-Q_{0}\equiv\delta
Q_{1}\tag{7d}\label{007d}%
\end{align}
and all these quantities are infinitesimal with $1/m$. So that we are entitled
to expand (\ref{005}) in $\delta Q_{CD}$ and keep only the first term in its
Taylor expansion, provided the resulting summation does not ruin the
infinitesimality. Thus from (\ref{005}) we obtain, the formal WT identity%
\begin{equation}
\sum_{CD}\frac{\partial^{2}F}{\partial Q_{AB}\partial Q_{CD}}\delta
Q_{CD}=\delta H_{AB}\tag{8}\label{008}%
\end{equation}
Note that this relationship has zero on its RHS for ($AB$) as in (\ref{007a}),
or a non-zero RHS in $\delta H$ for ($AB$) as in (\ref{007b}, \ref{007c},
\ref{007d}).

\section{Some notations:}

To write out in detail eq. (\ref{008}) we need to introduce some notation for%
\begin{equation}
\frac{\partial^{2}F}{\partial Q_{AB}\partial Q_{CD}}\equiv M^{AB;CD}\text{
.}\tag{9}\label{009}%
\end{equation}
Noting the \textit{overlaps} $A\cap B$%
\begin{align}
a\cap b  & =1\tag{10a}\label{010a}\\
\alpha\cap\beta & =2\tag{10b}\label{010b}\\
a\cap\alpha & =0\tag{10c}\label{010c}%
\end{align}
the matrix $M^{AB;CD}$ will be identified by its \textit{overlaps} $A\cap B$
and $C\cap D$ written\ as upper indices. To have a complete set of matrix
elements we need also to specify how many maximal \textit{cross-overlaps} we
have: $0$ if $AB\neq CD$, $1$ if $A=C$, or $B=C$, or $A=D$, or $B=D$, and $2$
if $A=C$, $B=D$ or $A=D$, $B=C$. This closeness index we write as a heavy
lower index. For example we have%
\begin{align}
M^{ab;ab}  & =M_{\mathbf{2}}^{1;1}\tag{11a}\label{011a}\\
M^{ab;ac}  & =M_{\mathbf{1}}^{1;1}\tag{11b}\label{011b}\\
M^{ab;cd}  & =M_{\mathbf{0}}^{1;1}\tag{11c}\label{011c}%
\end{align}
Alike for the $1\longleftrightarrow2$ exchange in the upper indices. Also%
\begin{align}
M^{a\alpha;a\alpha}  & =M_{\mathbf{2}}^{0;0}\tag{12a}\label{012a}\\
M^{a\alpha;b\beta}  & =M_{\mathbf{0}}^{0;0}\tag{12b}\label{012b}%
\end{align}
The only ambiguity left is to distinguish $M^{a\alpha;\alpha b}$ from
$M^{\alpha a;a\beta}$ which we write%
\begin{align}
M^{a\alpha;\alpha b}  & =M_{\mathbf{1}\left(  2\right)  }^{0;0}\tag{13a}%
\label{013a}\\
M^{\alpha a;a\beta}  & =M_{\mathbf{1}\left(  1\right)  }^{0;0}\tag{13b}%
\label{013b}%
\end{align}
exhibiting in parenthesis whether the repeated replica is roman ($1$) or greek
($2$).

\section{WT identity for $AB=a_{1}b$ :}

We carry out explicitly the $\sum_{CD}$ summation. We have%
\begin{multline}
\left(  M^{a_{1}b;a_{1}b}+\sum_{c}M^{a_{1}b;a_{1}c}+\sum_{b_{1}}%
M^{a_{1}b;b_{1}b}+\sum_{b_{1}c}M^{a_{1}b;b_{1}c}\right)  \left(  -\delta
Q_{1}\right) \tag{14}\label{014}\\
+\left(  \sum_{\alpha}M^{a_{1}b;a_{1}\alpha}+\sum_{\alpha b_{1}}%
M^{a_{1}b;b_{1}\alpha}\right)  \left(  \delta Q_{2}\right) \nonumber\\
+\left(  \sum_{\alpha_{1}}M^{a_{1}b;\alpha_{1}b}+\sum_{\alpha_{1}c}%
M^{a_{1}b;\alpha_{1}c}\right)  \left(  \delta Q_{1}\right) \nonumber\\
+\left(  \sum_{\alpha_{1}\beta}M^{a_{1}b;\alpha_{1}\beta}\right)  \left(
-\delta Q_{2}\right)  =-\delta H_{1}\nonumber
\end{multline}
Expliciting the summations and using the above notations (11)-(13) we get%
\begin{multline}
\left[  M_{\mathbf{2}}^{1;1}+\left(  m-p-1\right)  M_{\mathbf{1}}%
^{1;1}+\left(  p-1\right)  M_{\mathbf{1}}^{1;1}+\left(  p-1\right)  \left(
m-p-1\right)  M_{\mathbf{0}}^{1;1}\right]  \delta Q_{1}\tag{15}\label{015}\\
-\left[  \left(  n-m-p\right)  M_{\mathbf{1}\left(  1\right)  }^{1;0}+\left(
p-1\right)  \left(  n-m-p\right)  M_{\mathbf{0}}^{1;0}\right]  \delta
Q_{2}\nonumber\\
-\left[  pM_{\mathbf{1}\left(  1\right)  }^{1;0}+p\left(  m-p-1\right)
M_{\mathbf{0}}^{1;0}\right]  \delta Q_{1}\nonumber\\
+\left[  p\left(  n-m-p\right)  M_{\mathbf{0}}^{1;2}\right]  =\delta
H_{1}\nonumber
\end{multline}

The first observation is that if we keep the limits $n\rightarrow0$, followed
by $m\rightarrow\infty$ as taken in \cite{BM79}, the LHS of (\ref{015})
contains factors going to infinity and it is no longer justified to replace
eq. (\ref{005}) by the first term in its Taylor expansion eq. (\ref{008}). If
we want to get rid of the terms in $m$ in (\ref{015}) we can choose%
\begin{equation}
m\equiv n^{1/2}\mu\tag{16}\label{016}%
\end{equation}
and with $n\rightarrow0$ first, and then $\mu\rightarrow\infty$. Note that a
choice $m=n^{\alpha}\mu$, for $\alpha<1/2$ would sent $Q_{0}$ to infinity. As
for the choice $\alpha>1/2$ it would imply $Q_{1}=Q_{2}=Q_{0}=Q $, leaving no
room for the identities looked after. With the special choice $\alpha=1/2$, in
the limit $n=0$ we have (2) replaced by%
\begin{align}
Q_{1}  & =Q_{2}=Q\tag{17a}\label{017a}\\
Q_{0}  & =Q+\frac{Q}{2\mu^{2}}\tag{17b}\label{017b}%
\end{align}
and (7) by%
\begin{align}
\delta Q_{0}  & =0\tag{18a}\label{018a}\\
\delta Q_{1}  & =\delta Q_{2}=-\frac{Q}{2\mu^{2}}\equiv\delta Q\tag{18b}%
\label{018b}%
\end{align}
Thus for $\mu$ large we have an infinitesimal transform. (Note that with
(\ref{018a}) we did not bother to write terms in $\delta Q_{0}$ that occur in
eq. (\ref{014})).

With $n=0$ and with the choice of (\ref{016}) we now get%
\begin{equation}
\left[  M_{\mathbf{2}}^{1;1}-2M_{\mathbf{1}}^{1;1}+M_{\mathbf{0}}%
^{1;1}\right]  \delta Q+p^{2}\left[  -M_{\mathbf{0}}^{1;1}+\left(
2M_{\mathbf{0}}^{1;0}-M_{\mathbf{0}}^{1;2}\right)  \right]  \delta Q=\delta
H\tag{19}\label{019}%
\end{equation}
Here $p$, the number of replicas in the exchanged bunch can be any finite
integer, $p>1$. So clearly, if we wish an unambiguous answer, it would be
necessary that from other equations the coefficient of $p^{2}$ be set equal to zero.

\section{Related WT identities for $AB=a_{1}b_{1}$\newline and $AB=ab$ :}

Both identities have a vanishing RHS, $a_{1}b_{1}$ leads to $\delta H_{0}$
(vanishing as in (\ref{018a})) and $ab$ leads to zero (as in (\ref{007a})).
The calculation follows exactly the same line as in the previous section.
Spelling out the two equations obtained yields (i) the vanishing of the
coefficient of $p^{2}$ in (\ref{019}), giving the diagonal component
$M_{\mathbf{0}}^{1;1}$ in terms of the off-diagonal components $M_{\mathbf{0}%
}^{1;0}$, $M_{\mathbf{0}}^{1;2}$, (ii) the corresponding relationship for
$M_{\mathbf{1}}^{1;1}$ (see below).

\section{WT identities exhibited:}

With the above we can now express WT identities obtained for $A\cap B=1$ :

Replicon for overlap $1$ :%
\begin{equation}
M_{\mathbf{2}}^{1;1}-2M_{\mathbf{1}}^{1;1}+M_{\mathbf{0}}^{1;1}=\frac{\delta
H}{\delta Q}\tag{20a}\label{020a}%
\end{equation}

and%
\begin{align}
M_{\mathbf{0}}^{1;1}  & =2M_{\mathbf{0}}^{1;0}-M_{\mathbf{0}}^{1;2}%
\tag{20b}\label{020b}\\
M_{\mathbf{1}}^{1;1}  & =M_{\mathbf{1}}^{1;0}+M_{\mathbf{0}}^{1;0}%
-M_{\mathbf{0}}^{1;2}\tag{20c}\label{020c}%
\end{align}
In exactly the same way one gets corresponding equations for $A\cap B=2$%
\begin{equation}
M_{\mathbf{2}}^{2;2}-2M_{\mathbf{1}}^{2;2}+M_{\mathbf{0}}^{2;2}=\frac{\delta
H}{\delta Q}\tag{21a}\label{021a}%
\end{equation}%
\begin{align}
M_{\mathbf{0}}^{2;2}  & =2M_{\mathbf{0}}^{2;0}-M_{\mathbf{0}}^{1;2}%
\tag{21b}\label{021b}\\
M_{\mathbf{1}}^{2;2}  & =M_{\mathbf{1}}^{2;0}+M_{\mathbf{0}}^{2;0}%
-M_{\mathbf{0}}^{1;2}\tag{21c}\label{021c}%
\end{align}
Finally taking $A\cap B=0$, one gets%
\begin{equation}
M_{\mathbf{2}}^{0;0}-\left(  M_{\mathbf{1}\left(  1\right)  }^{0;0}%
+M_{\mathbf{1}\left(  2\right)  }^{0;0}\right)  +M_{\mathbf{0}}^{0;0}%
=\frac{\delta H}{\delta Q}\tag{22a}\label{022a}%
\end{equation}%
\begin{align}
M_{\mathbf{0}}^{0;0}  & =\frac{1}{2}\left(  M_{\mathbf{0}}^{1;0}%
+M_{\mathbf{0}}^{2;0}\right) \tag{22b}\label{022b}\\
M_{\mathbf{1}\left(  1\right)  }^{0;0}  & =M_{\mathbf{1}}^{1;0}+\frac{1}%
{2}\left(  M_{0}^{2;0}-M_{\mathbf{0}}^{1;0}\right) \tag{22c}\label{022c}\\
M_{\mathbf{1}\left(  2\right)  }^{0;0}  & =M_{\mathbf{1}}^{2;0}-\frac{1}%
{2}\left(  M_{\mathbf{0}}^{2;0}-M_{\mathbf{0}}^{1;0}\right) \tag{22d}%
\label{022d}%
\end{align}

\section{Effect of a magnetic field:}

At mean-field level it is easily verified that the equations of motion that
yield (2) are proportional to the equation giving the lowest eigenvalue
(Replicon) of the Hessian. That is, the WT identity for the Replicon zero mode
is trivially checked at the mean-field level. In the presence of an external
magnetic field (distinct from the unphysical conjugate fields $H_{AB}$) the
equations of motion yielding the order parameters is unchanged but for an
extra term $H$. This $H$ cannot appear in the Hessian, a second derivative
matrix, since in the Lagrangean it occurs in the linear term $H\sum_{AB}%
\phi_{AB}$, with $Q_{AB}=\left\langle \phi_{AB}\right\rangle $. Hence the
presence of an external magnetic field suppresses the Goldstone modes and
hence the transition, just like it occurs in $O\left(  N\right)  $ systems.

\section{A return on mean-field:}

Let us look back at eqs (2) which are only valid at the mean-field level.
Actually, in our derivation, we only have used a milder form of eqs (18). To
get the WT identities we only needed%
\begin{align}
\delta Q_{0}  & =0\tag{23a}\label{023a}\\
\delta Q  & \sim1/\mu^{2}\tag{23b}\label{023b}%
\end{align}
for the limits%
\begin{align}
n  & =0\tag{24a}\label{024a}\\
\frac{1}{\mu^{2}}  & \lll1\text{ .}\tag{24b}\label{024b}%
\end{align}
Is (23) valid to all orders beyond mean-field? This is easily checked to all
orders in the paramagnetic phase. Going back to (\ref{003},\ref{004}), we have
the order parameter $Q_{AB}$ defined by%
\begin{equation}
Q_{AB}=\partial W\left\{  H\right\}  /\partial H_{AB}\tag{25}\label{025}%
\end{equation}
that is by%
\begin{align}
Q_{AB}  & =\frac{1}{\mathcal{N}}\int\prod\limits_{\left(  CD\right)  }%
D\phi_{CD}\ \phi_{AB}\ \exp\left\{  \mathcal{L}\left\{  \phi\right\}
+\sum_{\left(  CD\right)  }H_{CD}\phi_{CD}\right\}  \tag{26a}\label{026a}\\
\mathcal{N}  & =\int\prod\limits_{\left(  CD\right)  }D\phi_{CD}\ \exp\left\{
\mathcal{L}\left\{  \phi\right\}  +\sum_{\left(  CD\right)  }H_{CD}\phi
_{CD}\right\}  \tag{26b}\label{026b}%
\end{align}
Here $\mathcal{L}\left\{  \phi\right\}  $ is the cubic BM Lagrangean, where
the fields $\phi_{AB}\left(  i\right)  $ are coupled via $\frac{w}{6}\sum
_{i}\sum_{ABC}\phi_{AB}\left(  i\right)  \phi_{BC}\left(  i\right)  \phi
_{CA}\left(  i\right)  $. We have everywhere omitted the space (site)
dependence since, in the end, the external source $H_{AB}\left(  i\right)  $
is always taken as $H_{AB}$, site independent.

Consider then the perturbation expansion of (25, 26) giving $Q_{AB}$ as a
power series of $H_{CD}$. If we choose $H_{CD}=H$, we then have $Q_{AB}=Q$ and
we write it as%
\begin{equation}
Q=f\left(  H\right)  H\text{ .}\tag{27}\label{027}%
\end{equation}
If we choose now $H_{ab}=H_{\alpha\beta}=H$ and $H_{a\beta}=H_{0}$, one then
has (no $H_{0}$ dependence when $n$ and $m$ vanish)%
\begin{equation}
Q_{ab}=Q_{\alpha\beta}\equiv Q=f\left(  H\right)  H\text{ .}\tag{28}%
\label{028}%
\end{equation}
One also has (because $W\left(  H,H_{0}\right)  $ can only be even in $H_{0}$)%
\begin{equation}
Q_{a\beta}\equiv Q_{0}=g\left(  H;H_{0}\right)  H_{0}\tag{29}\label{029}%
\end{equation}
and from (\ref{027}) when $H_{0}=H$,%
\begin{equation}
g\left(  H;H\right)  =f\left(  H\right)  \text{ .}\tag{30}\label{030}%
\end{equation}
Hence for $H_{0}-H\sim1/\mu^{2}$, we have%
\begin{equation}
Q_{0}=\left[  f\left(  H\right)  +\frac{\partial g\left(  H;H\right)
}{\partial H_{0}}\left(  H_{0}-H\right)  +\ldots\right]  H_{0}\tag{31}%
\label{031}%
\end{equation}
and%
\begin{equation}
Q-Q_{0}=\left[  f\left(  H\right)  +H\frac{\partial g\left(  H;H\right)
}{\partial H_{0}}\right]  \left(  H-H_{0}\right)  +\mathcal{O}\left(
H-H_{0}\right)  ^{2}\text{ .}\tag{32}\label{032}%
\end{equation}
That is, under the limits (24), one gets%
\begin{equation}
\delta Q\sim\delta H\text{ ,}\tag{33}\label{033}%
\end{equation}
thus justifying (23).

\section{Comments and Conclusion:}

We have thus exhibited the Goldstone behavior for the three Replicon modes
(20a, 21a, 22a), these modes acquiring a mass proportional to the (unphysical)
conjugate field $\delta H$, with in the end $\delta H=0$. A detailed
examination of the Hessian matrix components shows \cite{CrisantiDominicis}
besides that both the anomalous and longitudinal modes with zero overlap (as
in (22)) also remain massless.

Altogether we have 10 relationships (for 15 components). Note that with the
five off-diagonal components one builds the seven diagonal terms
$M_{\mathbf{0}}^{i;i}$, $M_{\mathbf{1}}^{i;i}$ with $i=1$, $2$, $0$ as in
(20bc), (21bc), (22bcd). The last three diagonal terms $M_{\mathbf{2}}^{i;i}$
(the one that contain the kinetic term for non-zero value of the momentum)
complete the setup.

To conclude we have given the right to exist to a spin glass theory whose
starting point is formally identical to Bray and Moore two-packets theory but
with the crucially different limits for the parameters $n$, $m$, namely with%
\[
m\equiv n^{1/2}\mu\text{\qquad(\ref{016}).}%
\]

With the limit $n=0$, $\mu\rightarrow\infty$ we have then derived ten
relationships between the fifteen components of the two-point (one particle
irreducible) functions. Relationships between three-point functions could be
derived in the same way all these relationships being
\textit{non-perturbative}.

This new Lagrangean is a good candidate to describe \textquotedblleft
droplet\textquotedblright\ aspects of the Ising spin glass theory. It raises
many questions and enjoys the following properties:\newline(i) As in BM, the
order parameter is, in the end, Replica Symmetric, a disguised
ferromagnet.\newline(ii) Its associated free-energy is probably worse (lower)
than the Parisi free-energy in the vicinity of the upper critical dimension.
What would be the effect of dimension (that enters via loops) and would there
be a critical dimension below which the \textquotedblleft
droplet\textquotedblright\ description would prevail is a crucial question to
investigate.\newline(iii) Its correlation functions enjoy several Goldstone
modes. These modes become massive in the presence of an external magnetic
field. They interact through cubic coupling. Thus their effective coupling
cannot vanish in the infra-red like is the case for $O\left(  N\right)  $
systems. It is thus expected that the $1/p^{2}$ behavior of the Goldstone
modes will develop anomalies. Large distance anomal behavior of correlation
functions (computed for example in $6-D$ dimension) will then have to be
confronted with numerically obtained droplet exponents.\vspace{1cm}

\noindent{\small The author is thanking E. Brezin, A. Crisanti and
T.\ Temesvari for useful discussions.}

\end{document}